\documentclass[aps, pre,showpacs,twocolumn,showemail]{revtex4}
\usepackage{bbm}
\usepackage{mathrsfs}
\usepackage[sans]{dsfont}
\usepackage{epsfig,psfrag}
\usepackage{amsmath,amsfonts,amssymb}
\usepackage[usenames]{color}
\usepackage[dvips, bookmarks, colorlinks=true, plainpages = false, citecolor = blue, linkcolor = blue, urlcolor = blue, filecolor = blue]{hyperref}
\def\be{\begin{equation}}
\def\ee{\end{equation}}
\def\bea{\begin{eqnarray}}
\def\eea{\end{eqnarray}}

\begin{document}

\preprint{draft}

\title{Classification of (2+1)$-$Dimensional Growing Surfaces Using Schramm$-$Loewner Evolution}
\author{A.A. Saberi $^1$}\email{a$_$saberi@ipm.ir}\author{H. Dashti-Naserabadi $^2$}\author{S. Rouhani $^2$}
\address {$^1$School of Physics, Institute for Research in Fundamental
Sciences (IPM), P.O.Box 19395-5531, Tehran, Iran \\$^2$ Department
of Physics, Sharif University of Technology, P.O. Box 11155-9161,
Tehran, Iran}

\begin{abstract}
Statistical behavior and scaling properties of iso-height lines in
three different saturated two-dimensional grown surfaces with
controversial universality classes are investigated using ideas from
Schramm-Loewner evolution (SLE$_\kappa$). We present some evidence
that the iso-height lines in the ballistic deposition (BD), Eden and
restricted solid-on-solid (RSOS) models have conformally invariant
properties all in the same universality class as the self-avoiding
random walk (SAW), equivalently SLE$_{8/3}$. This leads to the
conclusion that all these discrete growth models fall into the same
universality class as the Kardar-Parisi-Zhang (KPZ) equation in two
dimensions.
\end{abstract}


\maketitle

Nonequilibrium growth processes exhibit nontrivial scaling behavior
which are often characterized and classified by three exponents, the
roughness exponent $\alpha$, the dynamical exponent $z$ and the
growth exponent $\beta$~\cite{EW,stanley,Kardar,Meakin,HZ,Krug}.
Analytic results for the values of these exponents are scarce and
one has to depend on numerical analysis. In some cases, namely the
two-dimensional (2D) Kardar-Parisi-Zhang (KPZ) equation, numerical
results are not definitive and ambiguities remain. On the other hand
it is by no means clear that this set of exponents is exhaustive and
other characterizing exponents may exist. In this paper, we propose
a new method for analysis of two dimensional rough surfaces based on
Schramm-Loewner evolution \cite{schramm}. In this method iso-height
lines in the saturated regime are analyzed as random simple paths in
which no self-crossing occurs. This leads to an extra characteristic
for a grown surface, namely $\kappa$, the diffusivity coefficient of
SLE. Our approach arrives at a sharp conclusion on the universality
class of growing surfaces in two dimensions.

The discrete ballistic deposition (BD)~\cite{BD}, Eden, and
restricted solid-on-solid (RSOS) models~\cite{RSOS} (for a review
and definitions of these models see \cite{stanley}), are believed to
be in the same universality class as the KPZ equation~\cite{Kardar}
which describes a non-conserved growth.
\\ The evidence that two models belong to the same
universality class can be given in many ways. One of the most direct
approaches, as pointed out in~\cite{Schwartz}, is to show that the
two models correspond to the same fixed point system. In this
direction and based on master-equation approach, the KPZ equation
has been exactly derived for the RSOS model~\cite{Park}, which
indicates that these two models belong to the same universality
class.\\However, the story for the BD model is more controversial. A
continuum equation is derived from the BD microscopic rules
in~\cite{Schwartz} which deviates from the KPZ equation (the model
which is considered in~\cite{Schwartz} is the next-nearest-neighbor
(NNN) BD model, slightly different from the nearest-neighbor (NN) BD
model in our present paper). Despite this deviation, the symmetry
arguments suggest that the 1D BD system is in the same universality
class of the KPZ equation while for the 2D case, the absence of the
rotational symmetry in the derived continuum equation violates the
\emph{a priori} reason for them belonging to the same universality
class. An exact lattice Langevin equation for the BD model has been
derived in~\cite{Vvedensky}, whose continuum limit is shown to be
dominated by the KPZ equation. Although for a 1D substrate the
solution of the exact lattice Langevin equation yields the KPZ
scaling exponents, but for a 2D substrate its scaling exponents are
again different from those obtained from
simulations~\cite{Vvedensky}.
\\Another way to determine the universality class of a rough
surface is to compute its exponents $\alpha$, $\beta$ and $z$.
Numerical results are consistent with the proposition that RSOS and
KPZ models belong to the same universality class, but the situation
is more controversial when considering the BD and Eden models.

For the KPZ equation in $d=1$, the exact values $\alpha=1/2$ and
$\beta=1/3$ are known~\cite{Kardar}. The estimated values obtained
by various numerical works on BD in $d=1$ for roughness exponent and
growth exponent range from $\alpha= 0.42$ to $0.506$ and $\beta=
0.3$ to $0.339$~\cite{FV-85,Meakin-86,Baiod,Ko,Reis-01}. Among the
results, those obtained by Reis~\cite{Reis-01} are close enough to
the exact KPZ values. In $d=2$, there is no exact computation of the
exponents for the KPZ system, nevertheless, various numerical and
theoretical approaches have been applied to measure the exponents.
The simulations based on direct numerical integration of the KPZ
equation in 2D give $\alpha=0.37$ to $0.4$~\cite{Amar-90, Reis2},
and the values obtained by various theoretical methods range from
$\alpha=0.29$ to $0.4$~\cite{Katzav,Canet,Moore,Lassig}. Among the
theoretical approaches, application of the mode-coupling
approximation for the KPZ equation in 2D \cite{Moore} yielded
$\alpha\simeq0.38$, in good agreement with the values found from
simulations. However, the result $\alpha\simeq0.29$ obtained by the
self-consistent expansion for 2D KPZ equation \cite{Katzav} displays
a discrepancy with the results of simulations. \\The diversity of
the obtained values of various simulations for 2D BD model ranging
from $0.26$ to $0.38$ for $\alpha$ and $0.21$ to $0.24$ for
$\beta$~\cite{Meakin-86,Baiod,Ko,Reis-01,Family-90,Reis-04}, does
not indeed provide a convincing evidence that it belongs to the same
universality class of the KPZ model. The same story holds for the
Eden model with scattered reported results for $\alpha$ ranging from
$0.20$ to $0.39$ \cite{stanley} in (2+1) dimensions.

We have carried out extensive simulations of the RSOS, NN-BD, and
Eden models to estimate the values of the three exponents $\alpha$,
$z$ and $\beta$ in (2+1) dimensions. In some cases our results take
values out of the above mentioned ranges (see table \ref{Tab1}).
Although the two well-known scaling relations $\beta=\alpha/z$ and
$\alpha+z=2$~\cite{Fisher-91} are obeyed, within the statistical
errors. Nevertheless, due to the significant difference between the
exponent values of these three models, it is not possible to
conclude that they belong to the same universality class.

\begin{table}[lt]
\begin{center}
\caption{ Scaling exponents for BD, Eden, RSOS and KPZ
\cite{saberi1} models in two dimensions obtained by our simulations.
With the exception of KPZ, the averages were taken over $10^3$
independent simulation runs for different square substrates of size
$50\leq L\leq700$.} \label{Tab1}
\begin{tabular}{ccccccc}\hline\hline
Model & $\hspace{1.4cm}$ &$\alpha$& $\hspace{0.9cm}$ &$z$ &
$\hspace{0.9cm}$ & $\beta$
\\\hline

BD   &  & $0.28(2)  $   &  & $1.70(5)$ &  &  $0.15(1) $
\\
Eden &  & $0.36(2)$   &  & $1.65(5)$ &  &  $0.205(15) $
\\
RSOS &  & $0.393(10)$   &  & $1.58(3)$ &  &  $0.240(5)$
\\
KPZ &  & $0.37(1)$   &  & $1.61(3)$ &  &  $0.23(1)$
\\
\hline\hline
\end{tabular}
\end{center}
\end{table}

A new tool for study of domain walls in critical systems is the
theory of Schramm-Loewner Evolution (or SLE$_\kappa$) \cite{schramm}
(for a review see \cite{SLE}). The diffusivity constant $\kappa$
determines the critical exponents hence the universality class of
the system in question. Some authors have argued that SLE may be
applied to turbulence \cite{Bernard} and surface growth phenomena
\cite{WO3,saberi1,saberi2} as well. Recently, we have reported some
evidence of conformal invariance in the statistical properties of
iso-height lines in the saturated growth models including an
experimentally grown $WO_3$ surface \cite{WO3}, and numerical study
of the KPZ equation done by direct integration of the discretized
KPZ equation in (2+1) dimensions \cite{saberi1,saberi2}. The mere
observation of scale invariance in a 2D physical system does not
necessarily imply conformal invariance \cite{Polchinski,Riva}.
However arguments pointing to conformal invariance of 2D growing
surfaces have been attempted \cite{Moriconi}. This suggests the need
for more stringent tests of conformal invariance in such systems.

In this paper, we report the results of extensive simulations on
three different discrete growth models i.e., RSOS, NN-BD and Eden
models. We present evidence that iso-height lines on saturated
surfaces are SLE curves of the same diffusivity $\kappa=8/3$
indicating that in this spirit and within statistical errors, all
these models belong to the KPZ universality class.

The growth simulations were undertaken on an anisotropic geometry
i.e., strips of size $L_x \times L_y$ with $L_y=L$, $L_x=3L$ and
$50\leq L\leq 10^3$ (for Eden model, due to the long CPU time, the
simulations were done up to size $L=600$). Periodic boundary
conditions were applied in both directions (Strip geometry was
chosen for congruence with dipolar SLE). We gathered a number of
$5\times10^3$ of saturated samples of each size for each of the
three RSOS, BD and Eden models. The samples were obtained from
$10^2$ independent simulation runs. During each simulation run, each
of the $50$ samples was selected after $10^3$ time steps (in units
of the number of lattice sites) after the saturation time.
\begin{figure}[h]\begin{center}
\includegraphics[scale=0.31]{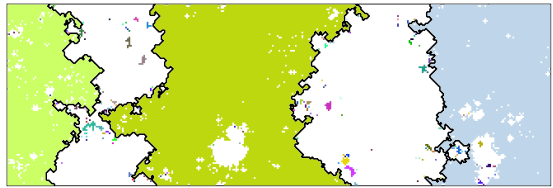}
\narrowtext\caption{\label{Fig0}(Color online) The positive-height
clusters shown in different colors, and the corresponding spanning
iso-height lines (solid lines). }\end{center}
\end{figure}

\begin{figure}[b]\begin{center}
\includegraphics[scale=1.6]{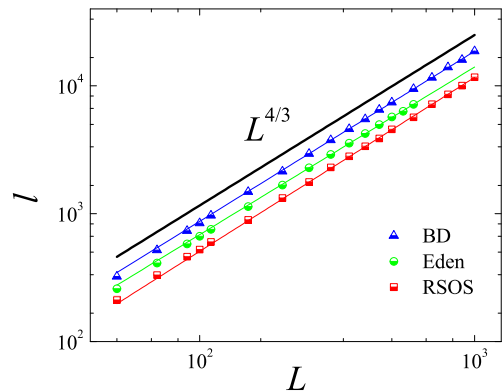}
\narrowtext\caption{\label{Fig1}(Color online) The average length
$l$ of a spanning iso-height line of the saturated growth models
versus the system size $L$, $l\sim L^{d_f}$. The solid line shows
the expected result for SAW. The error bars are almost the same size
as the symbols.}\end{center}
\end{figure}

For each height configuration $\{h_i\}$, a level cut is made at the
mean height, say $\bar{h}:=0$. Then the cluster heights were defined
as individual sets of connected sites with positive height which
were identified by using the Hoshen-Kopelman algorithm
\cite{Hoshen-Kopelman}. All spanning clusters along $y-$direction
were marked. For each spanning cluster, a walker algorithm was
applied to determine each of its perimeters which connects the lower
boundary to the upper one i.e., the spanning iso-height lines (Fig.
\ref{Fig0}). The iso-height lines were identified uniquely by using
the \emph{tie-breaking} rule on the square lattice, described in
\cite{Jstat}. Thereby, an ensemble of contour lines of fixed linear
size $L$ was obtained for each surface ensemble of different size.

We base our arguments on four different tests acertaining that the
iso-height lines are SLE: the fractal dimension, winding angle
statistics, left passage probability and direct SLE test.

i) \emph{Fractal dimension}. For conformally invariant curves the
fractal dimension is related to the diffusivity $\kappa$ by the
relation $d_f=1+\kappa/8$. The fractal dimension and the conjectured
value of diffusivity for SAW are known to be $d_f=4/3=1.3\bar{3}$
and $\kappa=8/3=2.6\bar{6}$.

\begin{figure}[b]\begin{center}
\includegraphics[scale=1.6]{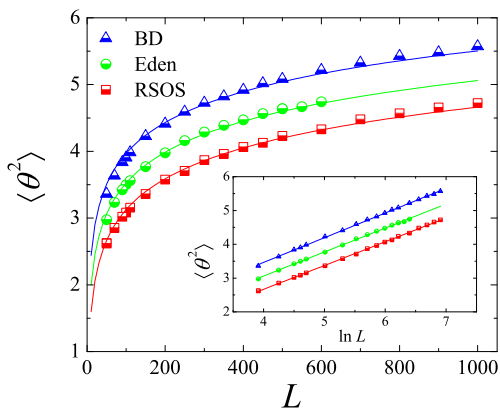}
\narrowtext\caption{\label{Fig2}(Color online) Variance of the
winding angle for the spanning iso-height lines of the saturated
growth models. The solid lines are set according to the Eq.
(\ref{winding}) for comparison, with appropriate obtained values of
$a$ for each model and $\kappa=8/3$. In the inset, the variance in
semilogarithmic coordinates.}\end{center}
\end{figure}

In Fig. \ref{Fig1} we show our computed results for the fractal
dimension of the iso-height lines for different models. The best fit
to our data in the whole range of the lattice sizes yields
$d_f=1.345(7), 1.330(5)$ and $1.335(4)$ for BD, Eden and RSOS
models, respectively. In all of our measurements for BD model, our
analysis on slightly larger system sizes $150\leq L\leq10^3$
corresponds to the same results as the Eden and RSOS models. For
example we find $d_f=1.337(5)$ for the BD model in this size range.
As shown in Fig. \ref{Fig1}, our results are well compatible with
the fractal dimension of SAW and SLE$_{8/3}$.

\begin{figure}[t]\begin{center}
\includegraphics[scale=1.48]{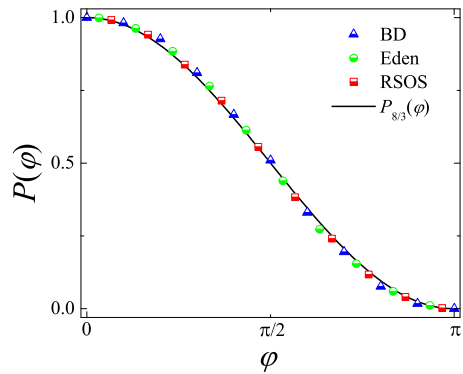}
\narrowtext\caption{\label{Fig3}(Color online) Left passage
probability computed for the spanning iso-height lines of different
models with $\rho=0.1 L$. The solid line shows the prediction of SLE
for $\kappa=8/3$. }\end{center}
\end{figure}
ii) \emph{Winding angle statistics}. The winding angle between two
end points of a finite SAW in two dimensions, is studied in
\cite{Duplantier_Saleur} using Coulomb gas methods. They found that
the winding angle is Gaussian distributed with a variance of $\sim
(8/g)\ln L$, where $L$ is the distance between the end points and
$g$ is Coulomb gas coupling parameter which is related to $\kappa$
by $g=4/\kappa$. They have also shown that the winding angle at a
single end point relative to the global average direction of the
curve is a Gaussian with variance of $(4/g)\ln L$. \\It is shown in
\cite{Wilson} that the variance in the winding
$\langle\theta^2\rangle$ at typical points along the curve is $1/4$
as large as the variance in the winding at the end points, i.e.,
\be\label{winding}\langle\theta^2\rangle=a+(\kappa/4)\ln L.\ee Using
the same definition as in \cite{Wilson}, we measured
$\langle\theta^2\rangle$ for different models, results are shown in
Fig. \ref{Fig2}. Data points compare well with Eq. (\ref{winding})
(solid lines), using $\kappa=8/3$ and a suitable value of the
parameter $a$ obtained from the best fit to data for each model. The
direct measurement of $\kappa$ also obtained from the best fits to
the data shown in the inset of the Fig. \ref{Fig2}. We find almost
the same value $\kappa/4= 0.700(20)$ for BD (again for larger
sizes), Eden and RSOS models, within the statistical error.

iii) \emph{Left passage probability}. The probability
$P_\kappa(\varphi)$ that an SLE$_\kappa$ curve, in the upper
half-plane, passes to the left of a given point at polar coordinates
($\rho$,$\varphi$), is computed by Schramm \cite{schramm2} \be
\label{Left_passsage}
P_{\kappa}(\varphi)=\frac{1}{2}+\frac{\Gamma\left(\frac{4}{\kappa}\right)}{\sqrt{\pi}\Gamma\left(\frac{8-\kappa}{2\kappa}\right)}
{}_2F_1\left(\frac{1}{2},\frac{4}{\kappa};\frac{3}{2};-\cot^2(\varphi)\right)\cot(\varphi),
\ee where ${}_2F_1$ is the hypergeometric function.\\As another
check, we measure this quantity (which should also hold for a
dipolar SLE near the starting point i.e., $\rho\ll L$ \cite{PRB})
for the contour lines of different growth models.\\As shown in Fig.
\ref{Fig3}, our results for the three models are again in good
agreement with the prediction for SLE$_{8/3}$.

\begin{figure}[t]\begin{center}
\includegraphics[scale=1.5]{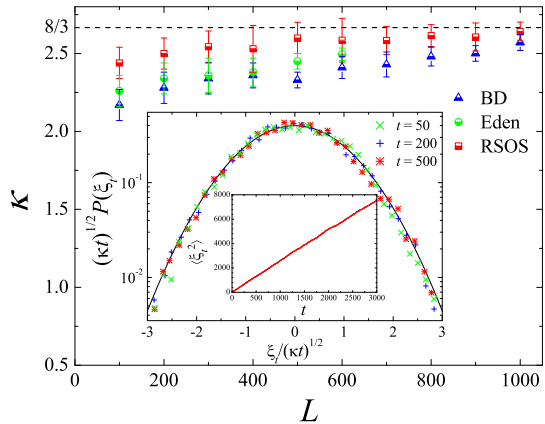}
\narrowtext\caption{\label{Fig4}(Color online) The diffusivity
$\kappa$ obtained for each contour ensemble versus their linear size
$L$. Inset: statistics of the driving function $\{\xi_t\}$ obtained
for the contour ensemble of the RSOS model with $L=10^3$.
}\end{center}
\end{figure}

iv) \emph{Direct SLE test}. Using a discrete Loewner evolution and
successive appropriate conformal maps, we extracted the Loewner
deriving function $\{\xi_t\}$ of each iso-height line represented by
the sequences of points $\{z_0, z_1,..., z_N\}$ in the complex
half-plane, with $z_0=(0,0)$. We use the function
$g_t(z)=\sqrt{(z-\xi_t)+4t}+\xi_t$, with $t=\frac{1}{4}\Im z_1^2$
and $\xi_t=\Re z_1$, to map all of the points except the first one
to a shortened renumbered sequence. After each recursive map the
first point in the sequence is swallowed and a sequence of
$\{\xi_t\}$ can be obtained for each iso-height line. We have also
checked the map appropriate for dipolar SLE \cite{PRB}, and found no
significant differences.\\We find that the statistics of the
deriving function for each curve ensemble of fixed linear size $L$,
converges to a Gaussian process with variance
$\langle\xi_t^2\rangle=\kappa t$. Finite size effects may be reduced
by looking at shorter segments of the curve e.g., when $10\%$ of the
total average length of the curves is mapped. An example is shown in
the inset of Fig. \ref{Fig4}, obtained for the contour ensemble of
the RSOS model with $L=10^3$. We observe that $\kappa$ shows a
slight dependence on system size, but reducing with $L$. We find
that the value of the diffusivity $\kappa$ for larger system sizes
approaches the expected value for SAW i.e., $\kappa=8/3$, for all
three models. As shown in Fig. \ref{Fig4}, for RSOS model this
convergence begins in rather smaller sizes but for Eden and BD
models larger system sizes are needed.

The more accurate results obtained here for the BD model is slightly
different from that reported in \cite{WO3}, this is due to the
considerable difference in the number of averaging samples and
reduced finite size effects.

Summing up, although the numerical values of three exponents
$\alpha$, $\beta$ and $z$ for BD, Eden, RSOS and KPZ models are
scattered, the numerical value found for $\kappa$ is sufficiently
sharp to suggest that these models all belong to the same
universality class. What remains is the inter-dependence of these
exponents and $\kappa$. The existence of two scaling relations
$\beta=\alpha/z$ and $\alpha+z=2$, guarantees that there is only one
independent exponent e.g., the roughness exponent $\alpha$. On the
other hand, this paper introduces a new characteristic value for a
rough surface: $\kappa$. How inter-dependent are these two? There is
a powerful scaling argument given in \cite{Kondev} which connects
$d_f$ of a contour line to $\alpha$ of the same surface, $d_f=2 -
x_l - \alpha/2$, where $x_l$ is the loop correlation exponent.
Although the exact value of $x_l=1/2$ is known only for the limiting
cases of $\alpha=0$ and $1$, but it is conjectured that its value is
super universal and is independent of $\alpha$ for Gaussian
surfaces. This leaves the case of 2D KPZ in ambiguity, since it does
not follow a Gaussian distribution. A simple minded value $x_l=1/2$
leads to $d_f=4/3$ giving $\alpha=1/3$ which although elegant
\cite{Canet} but is in conflict with numerical results. Our future
efforts will concentrate on revealing the nature of this
relationship.

\textbf{Acknowledgement.} A.A.S. acknowledges financial support from
INSF grant.

\end{document}